\documentclass[structabstract]{aa}  
\usepackage{graphicx}
\usepackage{sidecap}
\usepackage{txfonts}
\usepackage{natbib}

\newcommand{\lsim}{\mathrel{\hbox{\rlap{\lower.55ex \hbox{$\sim$}} \kern-.3em \raise.4ex \hbox{$<$}}}}

\usepackage{tabularx}
\newcolumntype{L}[1]{>{\raggedright\arraybackslash}p{#1}} % linksb殤dig mit Breitenangabe
\newcolumntype{C}[1]{>{\centering\arraybackslash}p{#1}} % zentriert mit Breitenangabe
\newcolumntype{R}[1]{>{\raggedleft\arraybackslash}p{#1}} % rechtsb殤dig mit Breitenangabe

\bibpunct{(}{)}{;}{a}{}{,} % to follow the A&A style

\begin{document}
   \title{Measurement of the earthshine polarization in the B, V, R, and I band
          as function of phase}

%   \subtitle{I. }

   \author{A. Bazzon\inst{1}, H.M. Schmid\inst{1}, D. Gisler\inst{2}}

   \institute{ETH Zurich, Institute of Astronomy,
              Wolfgang-Pauli-Str. 27, 8093 Zurich, Switzerland
              \and
   	     Kiepenheuer Institut f{\"u}r Sonnenphysik,
   		Sch{\"o}neckstr. 6, 79104 Freiburg, Germany\\
              \email{bazzon@astro.phys.ethz.ch} }

   \date{Received XX, 2013; accepted YY, 2013}

\authorrunning{A. Bazzon et al.}
%\titlerunning{Measurement of the earthshine polarization}

% \abstract{}{}{}{}{} 
% 5 {} token are mandatory
 
  \abstract
   {Earth-like, extra-solar planets may soon
become observable with upcoming high contrast 
polarimeters. Therefore, the characterization of the 
polarimetric properties of the planet Earth is important
for the interpretation of expected observations and
the planning of future instruments.}
   {Benchmark values for the polarization signal 
of the integrated light from the planet Earth 
in broad band filters are derived
from new polarimetric observations of the earthshine 
back-scattered from the Moon's dark side. }
   {The fractional polarization of the earthshine $p_{\rm es}$
is measured in the
B, V, R and I filters for Earth phase angles $\alpha$ between 30$^\circ$ 
and 110$^\circ$ with a new, specially designed wide 
field polarimeter. In the observations the light from the
bright lunar crescent is blocked with focal plane masks. 
Because the entire Moon is imaged, the earthshine observations
can be well corrected for the stray light from 
the bright lunar crescent and twilight. The phase
dependence of $p_{\rm es}$ is fitted by a function
$p_{\rm es}=q_{\rm max} \sin^2\alpha$. 
Depending on wavelength $\lambda$ and the lunar surface albedo $a$ the polarization of the back-scattered earthshine is significantly reduced. 
To determine the polarization of the planet Earth
we correct our earthshine measurements by a
polarization efficiency function for the lunar surface 
$\epsilon(\lambda ,a)$
derived from measurements of lunar samples from the literature.
}
   {The polarization of the earthshine
decreases towards longer wavelengths and
is about a factor 1.3 lower for the higher albedo highlands.
For mare regions the measured maximum polarization 
is about $q_{\rm max,B}=13$~\% 
for $\alpha=90^\circ$ (half moon) in the B band. The resulting fractional polarizations for
the planet Earth derived from our earthshine measurements and corrected by $\epsilon(\lambda ,a)$ are
24.6~\% for the B band, 19.1~\% for the V band, 13.5~\% for the
R band, and 8.3~\% for the I band. Together with literature
values for the spectral reflectivity we obtain a contrast $C_{\rm p}$
between the polarized flux of the planet Earth and
the (total) flux of the Sun with an
uncertainty of less than 20~\% and we find that the best phase to detect an Earth twin is around $\alpha = 65^\circ$.}
   {The obtained results provide a multi-wavelengths
and multi-phase set of benchmark values which are
useful for the assessment of different instrument and observing 
strategies for future high contrast polarimetry of extra-solar
planetary systems. The polarimetric models of Earth-like planets from Stam (2008) are in
qualitative agreement with our results but there are also significant
differences which might guide more detailed computations.}


   \maketitle
%
%=======================================================================

\section{Introduction}
This paper presents polarimetric observations of the 
earthshine on the Moon's dark side for a characterization 
of the integrated polarimetric properties of the planet Earth 
for the future investigation of Earth-like extra-solar planets.

With the rapid progress in observational techniques the detection 
of reflected light from terrestrial or even Earth-like extra-solar
planets may become possible in the near future with high-contrast imaging of
very nearby $(d \lsim 5pc)$ stars. Statistical studies based on the
radial velocity survey of stellar reflex motions due to 
low mass planets  \citep{Mayor2011}
or the planetary transit frequency 
of small planets by the KEPLER satellite 
\citep{Howard2012} 
indicate both 
that terrestrial planets could be present with high probability 
around every nearby star. The detection of a periodic RV-signal 
in $\alpha$ Cen B by
\citet{Dumusque2012}, 
which was 
attributed to a planet with a mass of $\approx 1 M_{\rm Earth}$, 
demonstrates that the very nearest stars are really excellent 
targets for the search of extra-solar planets.

The intensity contrast between a reflecting 
planet and the parent star is
\begin{equation}
 C_I(\alpha,\lambda)= f(\alpha,\lambda)(R_p/d_p)^2\,, 
\end{equation}
where $\alpha$ is the phase angle, $R_p$ the radius of the planet, 
$d_p$ its separation to the star, and $f(\alpha,\lambda)$ the phase
dependent reflectivity. Thus the contrast is high for small
separations $d_p$ and therefore the prospect for direct detection 
is particularly favorable for close-in planets $d_p \lsim 0.3$~AU 
around nearby stars for which such a small separation planet can still be
spatially resolved. However, detecting a faint signal from a 
reflecting planet at an angular separation of about 0.1 arcsec 
from a bright star is challenging and requires an instrument
with high spatial resolution and very high contrast capabilities
based on coronagraphy and some kind of differential imaging.
The upcoming planet finder instruments SPHERE 
\citep{Beuzit2008} and 
GPI \citep{Macintosh2012}
will provide both much improved performance 
for substantial progress in this direction. A particularly  
promising technique for the search of reflected light from 
planets around the nearest stars is differential polarimetric
imaging available with the SPHERE instrument. With sensitive 
polarimetry one can search for a polarized signal due to the 
scattered and therefore polarized light from the planet in the halo 
of the unpolarized light from the star
\citep[e.g.][]{Schmid2006}. 
The measureable
polarization contrast can be described similar to the intensity
contrast by 
\begin{equation}
C_p(\alpha,\lambda)= p(\alpha,\lambda) f(\alpha,\lambda)(R_p/d_p)^2\,,
\label{cpol}
\end{equation} 
where $p(\alpha,\lambda)$ is the
integrated fractional polarization. Therefore, the investigation of
$p(\alpha,\lambda)$ and the polarization flux 
\mbox{$p(\alpha,\lambda) \times f(\alpha,\lambda)$} 
of planet Earth is important for the planing of future observing projects on
extra-solar planetary systems and the interpretation of observational 
data. Up to now only very limited data are availabe for the
integrated polarization of the planet Earth. The oldest and still best
polarization phase curve of Earth originates from the 
observations of the earthshine by
\citet{Dollfus1957}.

He determined the earthshine polarization phase curve 
$p_{\rm es}(\alpha)$ with visual observations (i.e. in the V band) 
for Earth phase angles from $\alpha =22^\circ$, about 1.5 days after
new moon, to about $\alpha = 140^\circ$, between half and full moon. 
\citet{Dollfus1957} finds for dark regions (maria) on the Moon
a steady increase of the fractional polarization of the
earthshine from about $p_{\rm es}\approx 2$~\% around $\alpha=30^\circ$, 
to a maximum polarization of about $p_{\rm es}\approx 10$~\% 
for $\alpha\approx 100^\circ$, and a decline for larger
$\alpha$'s down to $p_{\rm es}\approx 4$~\% at $\alpha\approx 140^\circ$. 
\citet{Dollfus1957}
also finds a higher fractional polarization for
the back-scattered light for dark regions with surface albedo of
about $a=0.1$ than for bright regions with $a=0.2$. In addition
he notes a wavelength dependence in the fractional polarization
of the earthshine with higher values at shorter wavelength.
One should note that the back-scattering by the lunar surface 
introduces a depolarization of the earthshine. Thus, the 
fractional polarization of the light scattered by Earth is
higher by a factor of about 2 to 3 than the measured value from the
back-scattered earthshine. 

Space experiments did not provide much progress because full Earth
polarimetry was to our knowledge not taken or at least not published. 
Earth observing satellites with polarimetric capabilities took usually 
measurements of only small fractions of the Earth surface from which it is 
difficult to determine the net polarization for the entire planet. 
A result from the POLDER satellite was reported by 
\citet{Wolstencroft2005} 
who obtained fractional polarization values for $\alpha=90^{\circ}$ for three wavelengths and different
cloud coverages. For a typical value for the 
average cloud coverage of 55 \% 
they derive for the polarization of the planet Earth:
$p(443~{\rm nm})=22.6$~\%,  $p(670~{\rm nm})=8.6$~\%,  $p(865~{\rm nm})=7.3$~\%.

An interesting
new result on the Earth polarization from earthshine measurements is
the VLT spectro-polarimetry from 
\citet{Sterzik2012}
which show
narrow spectral features due to water, O$_2$, and O$_3$ 
absorptions in the Earth atmosphere and a rise of the fractional 
polarization towards the blue due to Rayleigh scattering.

Model calculations have been made for the fractional
polarization of the reflected light from Earth-like planets 
\citep{Stam2008} 
as well as the polarization produced by reflecting clouds
\citep[e.g.][]{Karalidi2011, Karalidi2012, Bailey2007}
or glint from ocean water surfaces 
\citep{Williams2008}. 
The models
provide an adequate description of the dominating scattering processes
and the signatures of different surface types. However, the overall
net polarization of Earth depends strongly on the not so well
known contributions of the different areas to the total signal. 
Therefore it is very desirable to have better observational
data which constrain between the various model options.   
 
Lunar earthshine observations are very attractive for the investigation
of the intensity and polarization of the reflected light of the
Earth because they provide the integrated scattered light signal 
from the whole planet Earth from the ground. However, for the
retrieval of the real level of scattered intensity $f_{\rm E}$ and
fractional polarization $p_{\rm E}$ of the Earth from the measured 
earthshine signals $f_{\rm es}$ and $p_{\rm es}$ 
one needs also to consider the back-scattering properties of 
the absorbing and depolarizing lunar surface.  

In addition, earthshine observations are very special because 
the Moon is a bright and large target for modern astronomical
instrumentation and because the contrast between the bright crescent 
and the dark side of the Moon is very high.
It is therefore not straightforward to disentangle
the earthshine from the disturbing contributions of the variable 
atmospheric (and instrumental) stray light from the bright moonshine 
and of the twilight. 

In this paper we describe new earthshine polarization measurements
taken with an imaging polarimeter specially designed for earthshine
observations. Our data cover Earth phase angles from $30^{\circ}$ to $110^{\circ}$
providing calibrated
$p_{\rm es}(\alpha)$ curves in the four broad-band filters B, V, R and
I for lunar maria and highlands which are corrected for the stray light
from the moonshine and the sky background. Section \ref{s:instrument} describes
our instrument and Sect. \ref{s:observations} our measurements while Sect. \ref{s:datareduction} discusses
the data reduction. The observational results are given in
Sect. \ref{s:results} and then we discuss in Sect. \ref{s:retro-reflection} our correction for the depolarizing
effect of the lunar surface. The final polarization phase curves $p_{\rm E}(\alpha,\lambda)$
for Earth and the derivation of the polarization flux $p_{\rm E} \times f_{\rm E}$
and the Earth-Sun polarization contrast $C_{\rm p}$ are given in Sect. \ref{s: earth polarization}.
The last Section \ref{s:conclusions} gives a summary and discusses the potential of
earthshine polarization measurements.

%=======================================================================  
  
\section{Instrumentation}\label{s:instrument}

\subsection{Instrument requirements}
        
        \begin{figure}
        \centering
        \includegraphics[width=9cm]{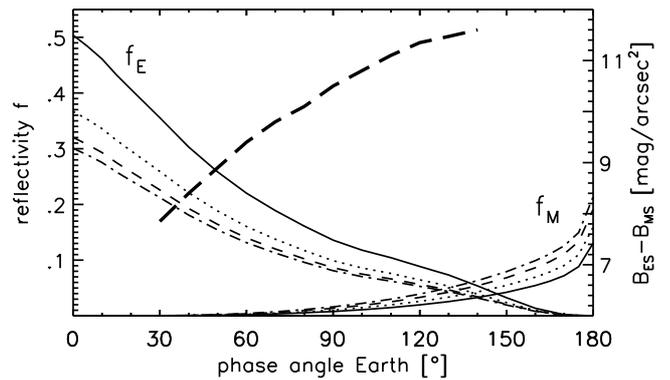}
                 \caption{Reflectivity of the Earth $f_{\rm E}(\alpha_{\rm E},\lambda)$
                 and the Moon $f_{\rm M}(\alpha_{\rm E},\lambda)$ for 
                 B (solid), V (dotted), R (dashed) and I (dash-dot). The thick dashed line is the difference of the surface brightness between 
                 earthshine $B_{\rm es}(\alpha_{\rm E},\lambda)$ and moonshine $B_{\rm M}(\alpha_{\rm E},\lambda)$ for the \mbox{400-700 nm} pass band. 
                 }
                 \label{evsm}    
        \end{figure}

       \begin{figure*}[!htb]
        \centering
        \includegraphics[width=18cm]{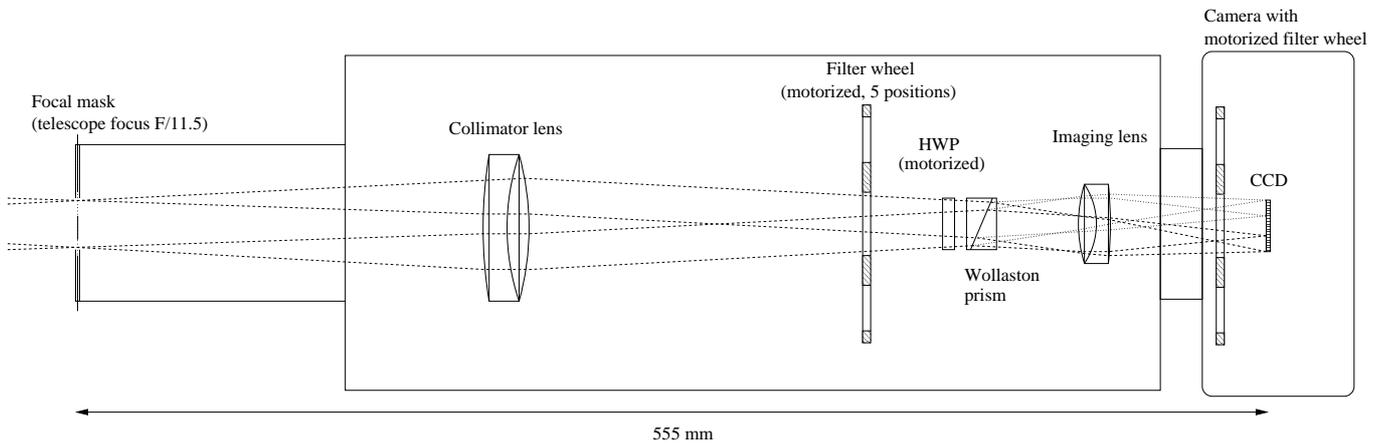}
                 \caption{Schematic overview of ESPOL with optical
        design and components drawn to scale. The focal mask is
        located in the focal plane of the 21cm telescope.
                 }
                 \label{setup}   
        \end{figure*}

The (surface) brightness 
of the lunar earthshine, the moonshine, and the contrast between
the bright and dark side of the Moon are described in the literature and summarized
in Fig. \ref{evsm}.
The reflectivity of the Moon 
$f_{\rm M}(\alpha_{M},\lambda)$ 
is from \citet{Kieffer2005} and plotted as function of the phase angle of the Earth
$\alpha_{\rm E}$ using $\alpha_{\rm E}=180^\circ - \alpha_{\rm M}$.
The reflectivity of the Earth $f_{\rm E}(\alpha_{E},\lambda)$ 
is from earthshine observations and model calculations 
by \citet{Palle2003} for the pass band 400-700 nm.
Due to the lack of multi-color earthshine observations the same shape 
for the reflectivity phase curve given by 
\citet{Palle2003}
is adopted for all colors and the curves for the different bands are just scaled
with the factors derived from earthshine spectra from \citet{Arnold2002} as described in Sect. \ref{flux contrast}.
Measurements of the surface brightness of the earthshine $B_{\rm es} (\alpha,\lambda)$ and the moonshine $B_{\rm M} (\alpha,\lambda)$ 
were derived from observations of fiducial patches of highland regions \citep[see also][]{Qiu2003} 
by 
\citet{Montanes2007} for the pass band \mbox{400-700 nm} and Earth phase angles between $\alpha_{\rm E} = 30^{\circ}-140^{\circ}$.
In Figure \ref{evsm} we adopt their waxing Moon results
and derive the mean daily surface
brightness contrast $B_{\rm es}(\lambda) - B_{\rm M}(\lambda)$ (thick dashed line) between the dark and the bright side of the Moon.
For $\alpha=90^\circ$ (half moon) the
surface brightness of the earthshine is 14.9 mag/arcsec$^{2}$. The difference between earthshine
and moonshine ranges from 8 - 12 mag/arcsec$^{2}$ between $\alpha_{\rm E} = 30^{\circ}-140^{\circ}$.

The surface brightness of the earthshine is highest for the
new moon phase and decreases with the Sun-Earth-Moon phase angle $\alpha$
(Fig. \ref{evsm}). Observations for small $\alpha$ near new moon require
daytime or twilight observations for which a correction
for the sky light is impossible or difficult. For $\alpha\approx
45^\circ$ an observational window of roughly 30 minutes with reasonably
dark sky conditions becomes available after sunset or before sunrise
for useful earthshine polarization measurements.

Observations during the night with 
much reduced sky background levels are possible for larger $\alpha$,
but the brightness of the moonshine due to the solar illumination increases rapidly 
(\mbox{Fig. \ref{evsm}}). At around $\alpha\approx 90^\circ$ the contrast between moonshine and earthshine
becomes greater than about 10$^{4}$ and the light scattering in the Earth atmosphere and the instrument becomes
more and more a problem for earthshine observations particularly in the red where the
moonshine is strong and the earthshine weak.
The light from the twilight sky and 
the moonshine are both strongly polarized $p>3~\%$ and this needs
to be considered for an accurate measurement of the earthshine 
polarization. The polarization of the moonshine is discussed in detail in Sect. \ref{s: moonshine}. 

\subsection{The earthshine polarimeter}\label{s:ESPOL}

The EarthShine POLarimeter (ESPOL) measuring concept takes the
background and stray light conditions for earthshine
observations into account. The instrument allows imaging polarimetry
of the entire Moon and the surrounding sky regions in order to
measure the polarization signal of the weak earthshine on top of 
the strong stray light from the moonshine 
and/or the light contribution from the sky. ESPOL includes in
addition exchangeable focal plane masks to block the light from
the bright moonshine. The blocking of the bright crescent is required to allow for 
integrations of a few seconds without heavy detector saturation.
 
ESPOL is a dual-beam imaging linear polarimeter based on the
rotating half-wave retarder plate and Wollaston polarization
beam splitter concept. A schematic overview of the instrument is given 
in Fig. \ref{setup}. The instrument includes a holder for exchangeable
focal masks with different Moon phase shapes to block the light 
from the bright lunar crescent in order to avoid 
heavy detector saturation. ESPOL uses a super-achromatic 
$\lambda / 2$ retarder plate on a motorized rotational stage for 
polarization beam switching and the selection of the $Q$ and $U$
polarization direction. The following Wollaston prism splits
the light into the ordinary $i_\parallel$ and extraordinary $i_\perp$ beams with polarization
perpendicular to each other. Both beams, each with a field of view
of $50'\times 40'$ 
are imaged on the same 3072 x 2048 pixel CCD 
detector with a pixel scale of 1.5 arcsec/pixel.  For our measurements we used
a pixel binning of $3\times3$ pixels which reduced the spatial resolution
to roughly 10 arcsec. 

Color or neutral 
density filters with a diameter of 5 cm or \mbox{2 inches} 
can be inserted into the five-position filter wheel
located in the collimated beam or into the camera 
filter wheel respectively. In order to optimally align the focal
masks to the orientation of the bright lunar crescent the 
mask holder can be rotated around the optical axis. In addition
the whole instrument can be rotated around the optical axis to
fix the zero-point of the polarization direction to any desired
orientation. 

ESPOL was built in-house for low costs using equipment for amateur
astronomers and standard polarimetric and optical components for
the wavelength range of 360-860 nm. The instrument is attached to 
an equatorially mounted 21 cm Dall-Kirkham Cassegrain telescope. 
As CCD system a thermo-electrically cooled SBIG-STL 6303E camera system 
with integrated filter wheel and shutter is used.                     
 
\begin{SCfigure}
        \centering
        \includegraphics[width=5.4cm]{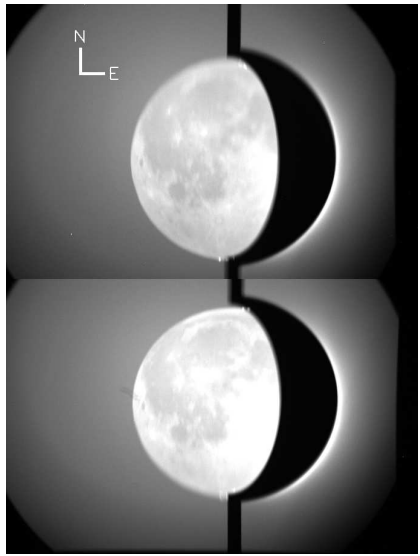}
                 \caption{ESPOL raw frame with the two polarization
        images $i_\parallel$ and $i_\perp$ of the
        earthshine on the Moon and the dark focal mask which blocks
        the light from the bright crescent.}
                 \label{raw}     
        \end{SCfigure}

Figure \ref{raw} 
illustrates the data format delivered by the CCD. 
Due to the large field of view and the use of simple optical
components the system shows quite some image distortion in
north-south direction where the Moon diameter between ordinary
and extraordinary beam differs by about 5\%. 
These distortions can be tolerated because we are not interested
in high spatial resolution but in the fractional polarization
of extended surface regions.

%=======================================================================

\section{Observations}\label{s:observations}

With ESPOL we measured the polarization of the earthshine 
for different phase angles and different wavelengths using a Bessell
B, V, R, I filter set \citep{Bessell2012}.  

To minimize read-out overheads and detector noise 
the CCD was operated with $3\times3$ pixel binning providing a
spatial resolution of about 10 arcsec, which is sufficiently high
for distinguishing mare and highland regions on the Moon. 

To minimize differential instrumental effects in the polarimetric
signal the measurements were performed in beam-exchange mode 
\citep{Tinbergen1996}. 
Only the linear Stokes components 
$Q/I$ and $U/I$ are measured. 

One polarimetric cycle consists of
two measurements with half-wave plate position $0^\circ$ and
$45^\circ$ for $Q/I$ and two measurements with position $22.5^\circ$ and
$-22.5^\circ$ for $U/I$. Typical integration times per exposure
were about 2-10 s per half-wave plate position so that one full cycle 
could be recorded in about 
one minute. This is sufficiently fast to avoid problems with
guiding drifts and strongly changing atmospheric conditions. 
To improve the S/N ratio typically a series of about 5 to 10 such datasets 
were recorded for each wavelength band during one observing night. 

ESPOL was rotated for all our measurements, including standard stars, 
into the orientation of the plane Sun-Earth-Moon so that the 
Stokes $+Q$ direction is a polarization perpendicular to this plane
and $-Q$ a polarization in this plane. The Stokes $\pm U$ directions
are $\pm 45^\circ$ with respect to this plane. The alignment was done by eye
by rotating the complete instrument until the crescent shaped focal mask completely blocked 
the moonlight which lead to an
alignment accuracy of about $\Delta\theta=\pm 2^\circ$.

For instrument monitoring and calibration additional polarimetric 
measurements of the moonshine and polarized/unpolarized standard stars 
as well as darks and twilight flatfield calibrations were recorded 
during each observing night. 
                
Our data were collected during two observing runs in March and 
October 2011. For the run in March the instrument was 
installed at the former Swiss Federal Observatory in Zurich 
at an altitude of 470 m above sea level. This 
run served mainly for first instrument testing and data with 
limited wavelength and phase coverage were obtained. Despite the 
non-optimal observing location in the heart of the city  
the data quality was good enough to be included in this study. 
For the second run we moved the instrument to the former Arosa 
Astrophysical Observatory at an altitude of 2050 m a.s.l. located 
in the Alps of eastern Switzerland. This site provides a much darker
sky and a much reduced level of light scattering in the Earth
atmosphere allowing measurements of the earthshine polarization for
larger phase angles. 
        
Both observing runs cover phase angles for the waxing Moon only. For
the March measurements the earthshine originates mainly from   
the Atlantic Ocean, the Pacific Ocean, and the american continent while
in  October the earthshine was due to reflected light from South
America and the Atlantic Ocean. \mbox{Table \ref{observations}} gives 
an overview of the observed phase angles of our measurements, the used filters 
and the number of polarization cycles for each filter. 

        \begin{table}
        \caption{Observing log. 
        The number of polarization cycles refer to the different
         filters taken for this date.}                                   
        \label{observations}      
        \centering          
        \begin{tabular}{C{19mm} C{10mm} C{13mm} C{18mm} }     
        \hline       
        observing date & phase      & filters   & \# pol. cycles \\
        \hline                    
        07.03.2011    & $31.5^{\,\circ}$       & V                     & 11          \\
        08.03.2011    & $42.5^{\,\circ}$       & B, V, R           &  3, 9, 2      \\
        11.03.2011    & $75.5^{\,\circ}$       & B, V, R            & 8, 19, 4    \\
        02.10.2011  & $73.0^{\,\circ}$        & B, V, R, I          & 1, 3, 4, 4 \\
        03.10.2011  & $85.5^{\,\circ}$        & B, V, R, I          & 5, 5, 4, 5            \\
        04.10.2011  & $98.0^{\,\circ}$        & B, V, R, I          & 16, 15, 21, 6       \\
        05.10.2011  & $109.5^{\,\circ}$      & B, V, R, I          &  8, 8, 10, 8     \\
        \hline                  
        \end{tabular}
        \end{table}

%=======================================================================

\section{Data reduction}\label{s:datareduction}

\subsection{Polarimetric reduction}

Figure \ref{raw} shows a typical ESPOL raw
frame with the ordinary $i_\parallel$ and extraordinary $i_\perp$
beams from the Wollaston showing the earthshine on the Moon in 
two opposite polarization directions. The 
bright crescent is blocked by the focal mask in order to suppress
stray light in the instrument and to avoid disturbing detector saturation. 
        
In the first data reduction step the raw images were dark subtracted
before the two opposite polarization images $i_\parallel$ and
$i_\perp$ for all half-wave plate orientations
($0^\circ,45^\circ,22.5^\circ$ and $-22.5^\circ$) were cut out and aligned.
Then the fractional Stokes parameter $Q/I$ images were calculated
according to the beam-exchange method described in
\citet{Tinbergen1996}:
\begin{equation} \label{beam-exchange calculation}
q = \frac{Q}{I} = \frac{R - 1}{R + 1} \hspace{0.5 cm} 
    with \hspace{0.5 cm} R^{2} =
\frac{i_{0,\parallel}/i_{0,\perp}}{i_{45,\parallel}/i_{45,\perp}},
\end{equation}
 where the first index of the image {\em i} refers to the $\lambda/2$
 retarder orientation and $\parallel$ and $\perp$ indicate the 
 two opposite polarization states from the ordinary and extraordinary
 beams of the Wollaston prism.         

The corresponding intensity images are calculated by
        \begin{equation}\label{intensity} 
        I_{Q} = 0.5 \cdot (i_{0,\parallel}+i_{0,\perp}+i_{45,\parallel}+i_{45,\perp})\,.
        \end{equation}
The polarization and intensity images for the \mbox{Stokes U}
 measurements are determined in the same way but using the frames
 taken with $+22.5^\circ$ and $-22.5^\circ$ retarder positions.
        
In the differential polarization measurements effects like the 
spatial variations of the system throughput, detector pixel-to-pixel
sensitivity differences and temporal changes between individual measurements
are compensated to first order with the used double ratio, without
any application of a flatfielding correction. Therefore, flatfielding was only applied on the intensity image
$I_{Q}$ described in Eq. (\ref{intensity}) using an intensity flatfield image produced in the same way. 

As described above there are some image distortions due to the
large field of view and the relatively simple optical setup.
These differential distortions between the
ordinary $i_\parallel$ and extraordinary $i_\perp$ beams
disappear almost entirely in the double ratio method because
images from both beams are in the nominator and denominator of
that ratio. This first order cancellation effect is not
present in the summed intensity images and leads to some spatial
smearing. Therefore the limb of the Moon is not sharp in the
intensity image but on the more relevant larger scales, 
i.e. for the identification of extended mare or highland regions, 
the image distortions are negligible. Nonetheless 
we have considered in our data analysis that small scale features
may be affected by image distortion effects and the 
associated alignment inaccuracies.   

The polarimetric properties of ESPOL were tested with observations
of zero polarization standard stars
$\beta$ Tau, $\beta$ UMa, $\gamma$ Boo, 
\citep{Turnshek1990} and Vega
\citep{Bhatt2000} which
show that the instrumental polarization is $\le 0.5\%$ in all 
filters. From the polarized standard stars
HD 21291, 9 Gem, $\phi$ Cas, \mbox{55 Cyg}
\citep{Hsu1982} 
we deduced a polarimetric efficiency above 98~\%
and checked the zero point of the polarization direction.
       
\begin{figure}
        \centering
        \includegraphics[width=8.5cm]{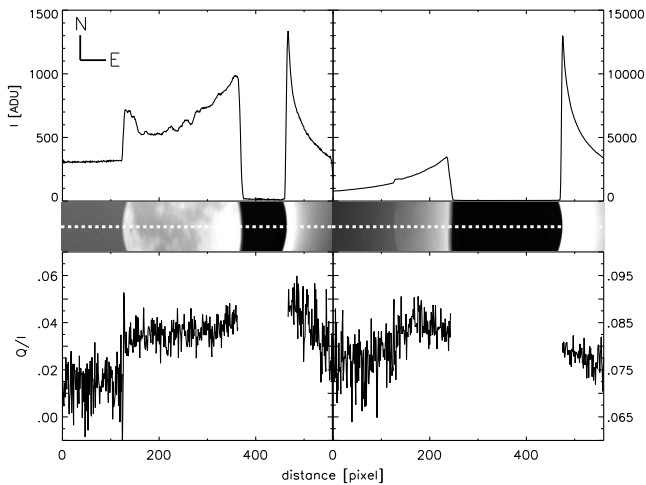}
                 \caption{Lunar west-east intensity (top) and 
        polarization (bottom) profiles in the B band for phase 42.5$^{\circ}$ and 98.0$^{\circ}$ with low (left) 
        and high (right) stray light contribution from the moonshine, respectively. 
        The panel in the middle shows the corresponding stripe of the
        intensity images. The profiles were extracted from 10 pixel wide
        regions as indicated by the dashed lines in the middle panel.}
                 \label{cuts}    
        \end{figure}

        \begin{figure}
        \centering
        \includegraphics[width=8.5cm]{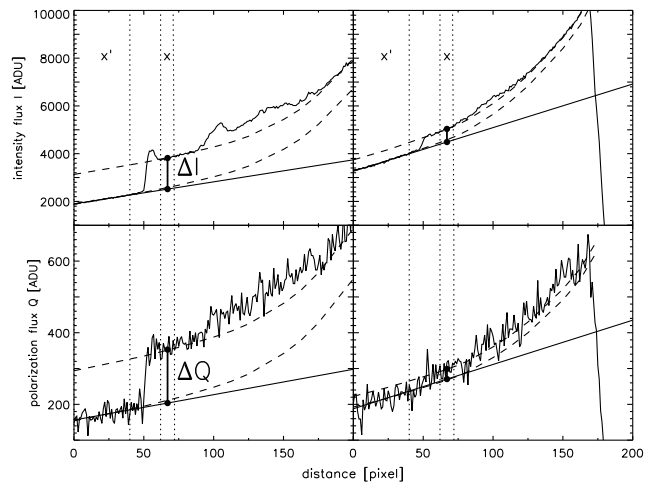}
                 \caption{Measuring the earthshine signals ${\Delta
        I},\,{\Delta Q}$ for relatively low (phase 73$^{\circ}$, region \#1, filter B)
        and high 
        (phase 98$^{\circ}$, region \#1, filter R) moonshine 
        levels  with the
        background $x'$ and measuring regions $x$ indicated. The 
        dashed lines illustrate the guessed level of
        the background (mainly stray light) and 
        background plus constant earthshine regions (reflected from maria).
        The full line is the linear extrapolation of the
        measured background from the $x'$ to the $x$ region.}
                \label{extract es}      
        \end{figure}

\subsection{Extracting the earthshine polarization}\label{s: extract earthshine}

We are interested in the measurement of the fractional polarization
of the earthshine $(Q/I)_{\rm es}$, 
which needs to be extracted from our data. Our observations show
the contributions of three intensity 
components 
\begin{displaymath}
I_{\rm tot}=I_{\rm es}+I_{\rm M}+I_{\rm sky}
\end{displaymath}
from the earthshine (es), the scattered light from the moonshine (M) 
and the sky (see Fig. \ref{cuts}). In our images it is rather
easy to distinguish these components, assuming that the sky is essentially
constant over the whole field of view. The location of the earthshine is well defined and 
its intensity lies in a restricted range between the intensities of dark
maria and bright highlands. The scattered light intensity from the moonshine 
has a more complex geometry. It is increasing rapidly towards the 
bright crescent which is covered in our data by the occulting mask.

The signatures of the three components can also be recognized in the
fractional polarization W-E cuts extracted from the $Q/I$ images 
shown in Figure \ref{cuts}. 
The B band observation for phase 42.5$^{\circ}$ 
shows a significant sky contribution from the twilight. The sky 
polarization is about
1.5 \% on the west side of the Moon and the earthshine plus sky
polarization is about 3.5 \%. The polarization of the moonshine
is slightly higher ($\sim$4.5 \%) as can be seen near the east side of the occulting 
mask where the scattered light of the moonshine dominates. For phase 
98$^{\circ}$ the
scattered moonshine with a polarization of about 8 \% dominates strongly.
The fractional polarization is just slightly enhanced at the position 
of the earthshine. The (Q/I)-images consist of the following
contributions
\begin{equation}
\left({Q\over I}\right)_{\rm tot} =  
        {Q_{\rm es}+Q_{\rm M}+Q_{\rm sky}\over I_{\rm tot}}\,.
\end{equation}  
Because of our definition of the $\pm Q$-directions 
perpendicular and parallel to the scattering plane the $U/I$ polarization is
essentially zero ($\approx \pm 0.5$ \%) and dominated by noise.

After some investigation we defined a procedure for the extraction of the 
fractional earthshine polarization $(Q/I)_{\rm es}$ which provides
also good results for large phase angles and the I filter 
for which the signal is weak and/or the stray light from the moonshine is very 
strong. For small phase angles the earthshine signal is strong and 
the measurement is easy. The basic idea is to measure the signal
of the earthshine on top of the "background signal" in the
I$_{\rm tot}$ frame and the Q frame where $Q=(Q/I)_{\rm tot} \cdot I_{\rm tot}$. 
The "background signals" (bg) are just the sum of the contributions
of the sky and the moonshine $I_{\rm bg} = I_{\rm sky} + I_{\rm M}$
and $Q_{\rm bg} = Q_{\rm sky} + Q_{\rm M}$. For this we 
extract radial cuts and extrapolate
the background signal from the region $x'$ outside 
to a location $x$ inside the lunar disk 
($I_{\rm bg}(x')$, $Q_{\rm bg}(x')\rightarrow 
I_{\rm bg}(x)$, $Q_{\rm bg}(x)$) where we measure the earthshine +  
background level. The final signal is then:
\begin{equation}  
 \left(\frac{Q}{I}\right)_{\rm es} = 
\frac{\Delta Q}{\Delta I}  = \frac{Q_{\rm bg+es}(x) -
  Q_{\rm bg}(x)}{I_{\rm bg+es}(x) - 
  I_{\rm bg}(x)} \,.
\end{equation} 

This procedure is illustrated in Figure \ref{extract es}, for
two cases. The first is a strong and clear earthshine signal typical for
phase angles $\alpha < 109^{\circ}$ in the B, V filters and phase angles $\alpha < 98^\circ$ in the R filter respectively.
The large majority of our data are of this kind.  
The other case is typical for phase angles $\alpha \ge 98^{\circ}$ 
in the R and I band filter
for which the stray light from the moonshine dominates strongly. 
The Q signal from the earthshine is still
above but close to the measuring limit. 
Also given are fits 
to the background, which consists in these cases mainly of
the moonshine plus a constant earthshine level fitted to the  
mare regions. The use of 
a linear extrapolation of the background in the $x'$ region 
for the background correction of the total earthshine plus background signal
measured at $x$ seems reasonable \citep[see also][]{Qiu2003, Hamdani2006}.

       \begin{figure}
        \centering
        \includegraphics[width=5.7cm,angle=90]{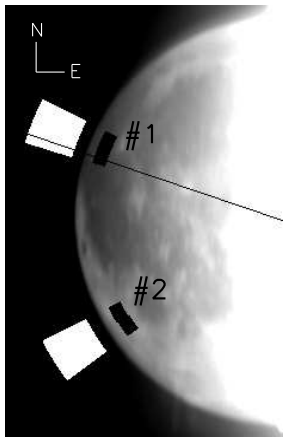}
                 \caption{Selected mare (\#1) and highland (\#2) fields used for
              the earthshine measurements together with their 
              background regions (white areas). }
                 \label{obs_areas}       
	\end{figure}

       \begin{figure*}
        \centering
        \includegraphics[width=17.6cm]{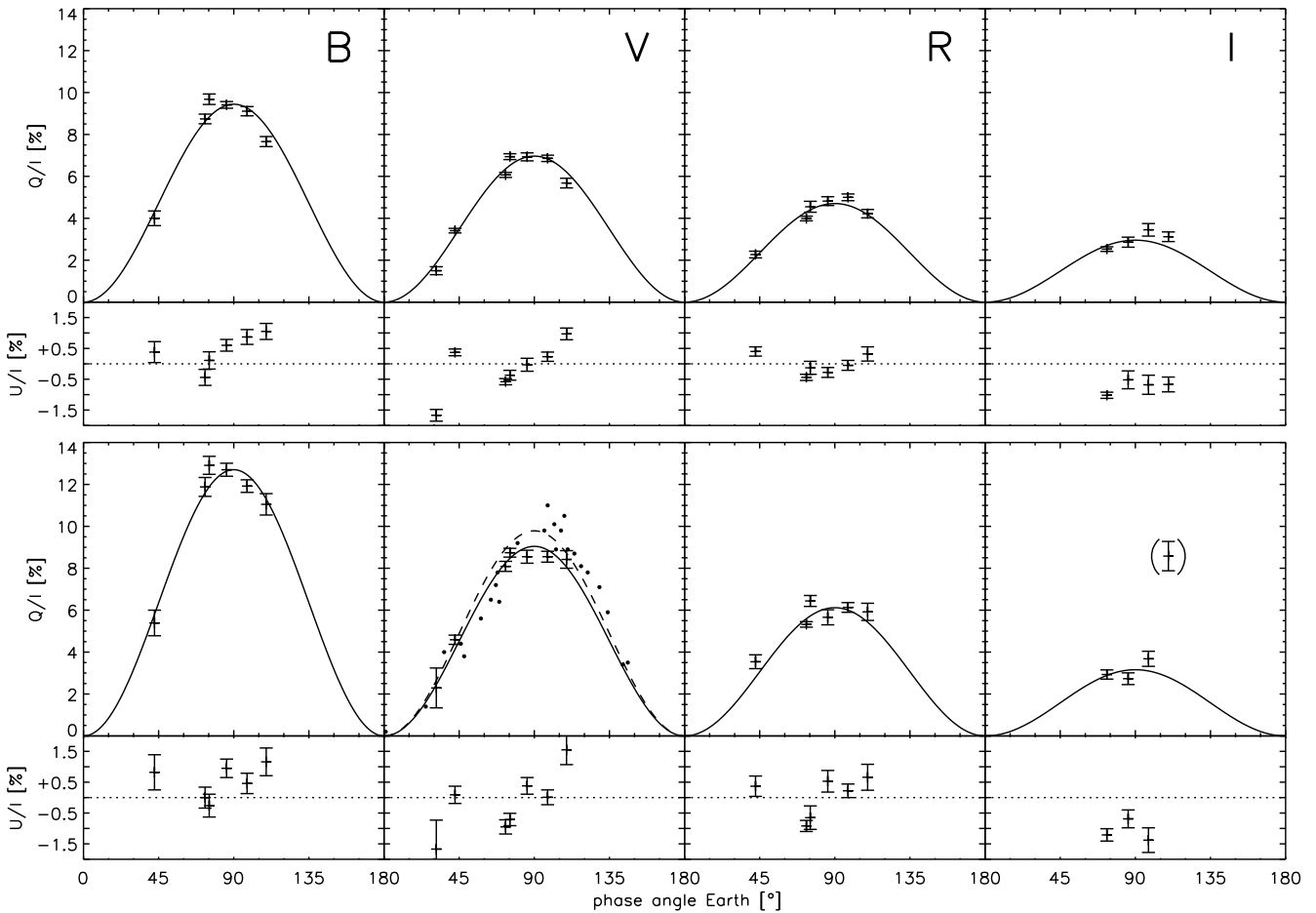}
                 \caption{Fractional polarization Q/I and U/I of the
        earthshine measured for highland (top) and mare regions
        (bottom) for the four different filters B, V, R and I (left to right).
        The solid curves are $q_{\rm max} \sin^{2}$ fits to the data. 
        The error bars give the statistical
        1$\sigma$ noise $\Delta_{\rm noise}$ of the data whereas the mare I band data at phase angle $109.5^{\circ}$ are
        additionally affected by a substantial systematic offset $\Delta_{\rm syst} > 0.5~\%$.    
        The dots in the V panel for the mare region indicate the measurements of 
        \citet{Dollfus1957} and a corresponding
        $q_{\rm max} \sin^{2}$ fit (dashed line) is also given.
                 }
                 \label{results}         
        \end{figure*}

        \begin{table*}
        \caption{Fractional polarization values $(Q/I)_{\rm es}$ 
        		     for the earthshine from the mare and highland regions of all our measurements and corresponding typical statistical 1$\sigma$ uncertainties $\Delta_{\rm noise}$. Also given is the fit parameter 
                        for $q_{\rm max} \sin^{2}(\alpha)$ derived in Sect. \ref{phase dependence} and the standard deviation of the data points from the fit $\sigma_{\rm d-f}$. 
                        The polarization efficiencies 
                        $\epsilon_{\#1}$ and $\epsilon_{\#2}$ are derived in Sect. \ref{s:retro-reflection} and the depolarization corrected values $q_{\rm max, corr}$ are given. 
                       }                               
        \label{tbl: results}      
        \centering          
         \begin{tabular}{C{16mm} | C{24mm} | C{11mm} C{11mm} C{11mm} C{11mm} | C{11mm} C{11mm} C{11mm} C{11mm}}     
            \hline    
        date &  phase      &   \multicolumn{4}{c | }{$Q/I$ (highland) [\%]}            & \multicolumn{4}{c}{$Q/I$ (mare) [\%]}  \\
        (2011)         & (Sun-Earth-Moon)                          &            B & V & R & I                                           &       B & V & R & I  \\
        \hline                    
        07.03  &   31.5$^{\circ}$       &     -          &      1.5    &          -             &    -             &       -            &       2.3       &          -           &      -               \\
        08.03  &   42.5$^{\circ}$       &  4.0        &      3.4    &         2.3         &     -            &       5.4         &       4.6       &       3.5        &       -               \\
        02.10  &   73.0$^{\circ}$       &  8.7        &       6.1   &         4.0         &    2.5         &      11.9      &       8.1       &        5.3       &        2.9          \\
       11.03  &   75.5$^{\circ}$       & 9.7      &      6.9    &          4.5         &    -             &     12.9        &      8.7        &      6.4         &      -               \\
        03.10  &   85.5$^{\circ}$      &   9.4     &       6.9   &           4.8        &   2.9         &      12.7       &        8.6      &        5.7       &        2.7            \\
        04.10  &   98.0$^{\circ}$      &   9.1       &       6.9    &          5.0        &    3.4         &      11.9       &       8.6       &       6.1        &        3.7            \\
        05.10  &  109.5$^{\circ}$     &    7.7      &        5.7  &            4.2       &    3.1        &       11.1      &         8.4    &          5.9     &         8.6$^a$          \\
        \hline
         \multicolumn{2}{l | }{statistical 1$\sigma$ uncertainty $\Delta_{\rm noise}$ [\%]} & 0.2 & 0.2  & 0.2  & 0.2  & 0.4  &  0.3   &  0.3   &   0.3   \\         
         \multicolumn{2}{l | }{fit parameter $q_{\rm max}$ [\%]} & 9.4 & 7.0  & 4.7  & 3.0  & 12.7  &  9.0   &  6.1   &   3.2   \\ 
         \multicolumn{2}{l | }{stddev of data from fit $\sigma_{\rm d-f}$ [\%]} & 0.22 & 0.13  & 0.10  & 0.21  & 0.24  &  0.13   &  0.22   &   -   \\ 
         \multicolumn{2}{l | }{polarization efficiency $\epsilon$  [\%]$^b$ }   &  36.6 & 34.3  & 32.7   & 30.7  & 54.3  &  50.8   &  48.5   &   45.6   \\ 
        \multicolumn{2}{l | }{$\epsilon$ corrected $q_{\rm max, corr}$  [\%]} & 25.8 & 20.3  & 14.4  & 9.6  & 23.4  &  17.8   &  12.6   &   6.9   \\
        \hline
        \multicolumn{10}{l}{\vspace{-0.2cm}}\\
        \multicolumn{10}{l}{$^a$value affected by systematic errors due to high stray light level (see text)}\\
        \multicolumn{10}{l}{$^b$the uncertainty of $\epsilon$ is mainly a systematic offset estimated to be $\Delta\epsilon = \pm 3~\%$ (see text)}  
        \end{tabular}
        \end{table*}

We have investigated more complex background/straylight correction
procedures, e.g. using 3-parameter exponential fits, but they didn't agree better than
the linear extrapolation. Important for the accuracy of 
the earthshine measuring process is the selection of areas close
to the western limb but not exactly at the limb because image 
alignment uncertainties of the polarimetric data reduction can create
disturbing spurious features at the limb. The limb is also not a
good measuring region because of the extreme incidence and
reflection angles (near $90^\circ$) with respect to the large scale 
surface normal which represents a situation which is not well explored
for its back-scattering properties.  
    
All our data show that the differences between the lunar dark mare
and bright highland regions are significant when determining the intensity
and polarization of the back-scattered earthshine. 
Therefore it is important to carry out separate measurements
for these two main lunar surface types. 
Because of the strong albedo dependence (see Sect. \ref{s:retro-reflection}) 
it is important that a chosen measurement field on the Moon does not have strong albedo variations. 
Under these terms we selected one mare field \#1 in the Oceanus Procellarum area
and one highland field \#2 between Mare Humorum and 
the Moon's limb as indicated in Figure \ref{obs_areas}. Both fields are close to the western limb far away
from the bright lunar side.  They are available for measurements at all phase angles $\alpha$ taking into account
increasing stray light and the lunar libration. 
Therefore, for both fields a consistent data reduction could be carried out.
  
For both fields $10$ radial
I and Q profiles separated by one degree were extracted and $\Delta I$
and $\Delta Q$ was determined as described above.  
Table \ref{tbl: results} gives the obtained $(Q/I)_{\rm es}$ polarization values 
from both fields
which are also plotted in Figure \ref{results} as phase curves together with the
statistical 1$\sigma$ error bars $\Delta_{\rm noise}$. 
    
As long as the S/N is sufficiently high 
the linear extrapolation method is robust. The total uncertainty 
for the obtained fractional polarization of the earthshine for
a highland or mare region at a particular date can be described by the statistical noise
plus a predominantly positive systematic offset
 $\Delta(Q/I)_{\rm es}= \Delta_{\rm syst}\pm \Delta_{\rm noise}$.

The statistical $1\sigma$ uncertainty $\Delta_{\rm noise}$ is small ($< 0.3~\%$).
This follows from
the scatter of the obtained values from different extraction
cuts of the same day and includes random noise, but also hard to
quantify systematic effects due to small image drifts on the detector,
changing stray light levels of the moonshine related to non-stable
atmospheric conditions, and perhaps other unidentified effects.    

For observations
with very small earthshine signals (i.e. at large phase angles and/or strong stray light of the moonshine) 
the linear extrapolation of the background 
introduces a systematic overestimate $\Delta_{\rm syst}$ of 
the result. This is because the moonshine dominated stray light background
increases with an upward curvature towards the illuminated 
crescent. For the B, V, and R measurements at phase angles $< 100^{\circ}$ this offset is negligible or small ($< 0.5~\%$).
However in the I band filter at phase angle $109.5^{\circ}$ the systematic offset is 
clearly dominating and the mare I band result for $109.5^{\circ}$ is no longer useful
(see Fig. \ref{results}). For this reason we disregard the mare I band result at $109.5^\circ$.

 \begin{figure}
        \centering
        \includegraphics[width=8cm]{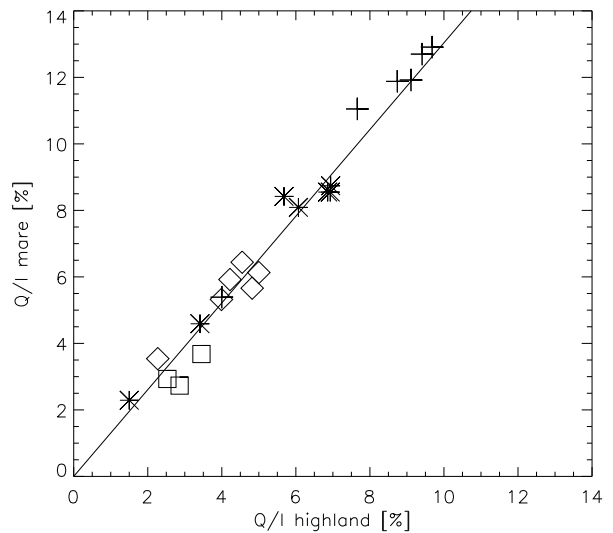}
                 \caption{Correlation between the fractional polarization 
                 for mare and highland regions measured simultaneously. The different symbols
                 indicate the colors 
                 B($+$), V($\ast$), R($\diamond$), and I($\square$). The line 
                 shows the derived proportionality factor $1.30\,\pm 0.01$.}
                 \label{ratio mh}        
        \end{figure}

%=======================================================================

\section{Earthshine polarization results}\label{s:results}

\subsection{Data}

The results for the fractional earthshine polarization $Q/I$
measured in the Bessell B, V, R and I bands are presented
in \mbox{Table \ref{tbl: results}} and Figure \ref{results}. 
The plots in Fig. \ref{results} also include the $U/I$ data points and the estimated
statistical 1$\sigma$ uncertainties of the individual data points $\Delta_{\rm noise}$.
The mare V band panel shows also the measurements by \citet{Dollfus1957}
which are in good agreement with our data.

Our earthshine data show a very good correlation between the 
polarization taken simultaneously 
for the highland and mare regions. Independent of color filter and phase
angle the polarization for the mare region is a factor of $1.30\pm0.01$ higher
than for the highland region as illustrated in
\mbox{Figure \ref{ratio mh}}.

\begin{figure*}
        \centering
        \includegraphics[width=18cm]{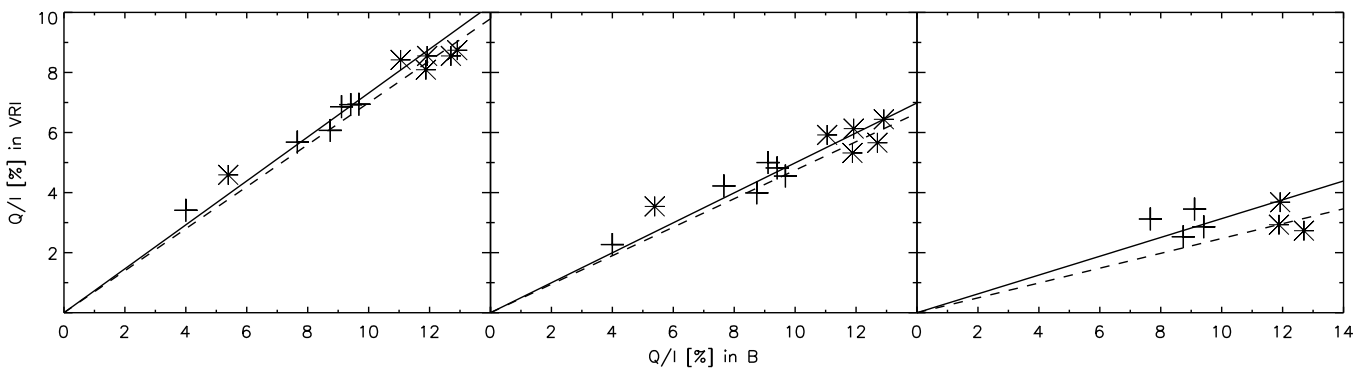}
                 \caption{
                 Fractional polarization of the earthshine reflected at highland ($+$) and mare ($\ast$)
                 regions in V, R and I band (left to right) with respect to the B band. The lines indicate linear fits
                 both for highland (solid) and mare (dashed) data separately. 
                 }
                 \label{ratio BF}        
        \end{figure*}

Good correlations are also found between different colors taken for the same observing date.    
When we plot the polarization $(Q/I)_{\rm es}$ in the V, R and I band versus the 
polarization in the B band (Fig. \ref{ratio BF}) 
we find that the ratios are independent of $\alpha_{\rm E}$. We get the ratios 
$0.72\pm0.02$, $0.49\pm0.02$ and $0.28\pm0.05$ for the ratios of the polarization between V and B band, 
R and B band, and I and B band respectively. Therefore, we conclude that to 
first order we can assume the same shape for the  polarization phase curve for
all wavelengths.

\subsection{Fits for the phase dependence}\label{phase dependence}

The phase dependence of the earthshine polarization looks symmetric
and can be well fitted with a simple $q_{\rm max}\sin^{2}(\alpha)$ curve. 
The model simulations by \citet{Stam2008} for Earth-like planets also support phase curves $q_{\rm max}\sin^{p}(\alpha)$.
She calculates polarization phase curves assuming a range of 
surface types (e.g. forest-covered areas with Lambertian reflection, dark ocean with 
specular reflection) and cloud coverages.
We find that the broad shape of her model phase curves can be well fitted by
curves $\sim q_{\rm max} sin^{p}(\alpha + \alpha_{0})$ with $p \approx 1.5-3$ and $\alpha_0 \approx 0^{\circ}-10^{\circ}$. 

Furthermore, she finds characteristic features at low 
phase angles due to the rainbow effect and negative polarization at large phase angles due to second
order scattering. We cannot assess the presence of such features because of 
the coarse phase sampling of our data.

Besides the $q_{\rm max} \sin^{2}(\alpha)$ curve we also tried functions with more free parameters to fit the data,
e.g. using a curve like $q_{\rm max} \sin^{p}(\alpha+\alpha_0)$
and varying the exponent $p$ between
values of 1.5-3 and by introducing a phase shift $\alpha_0$. However, such fits provide
not a significantly better match to the data. 
Because our data cover predominantly phase angles around quadrature the
shape of the phase curve is not very well constrained. 

The derived $q_{\rm max}$ fit parameters for the different phase curves are given in Table \ref{tbl: results}
together with the standard deviation of the data points from the fit $\sigma_{\rm d-f}$. For $Q/I$ the typical $\sigma_{\rm d-f}$ is $\approx 0.2~\%$ in good
agreement with the typical 1$\sigma$ uncertainty of the individual data points $\Delta_{\rm noise}$. The standard deviation of the derived
$U/I$ values from the expected zero-value is only slightly higher and typically \mbox{$\approx 0.3~\%$} indicating that the instrument alignment with respect to the
Sun-Earth-Moon plane was excellent (see Sect. \ref{s:observations}). Note that the $U$ signal is at the level of the measurement noise $|U|\approx\Delta_{\rm noise}(U)$. 
Therefore one should not use the normalized 
total polarization ${\rm p}=\sqrt{(Q/I)^2+(U/I)^2}$ because the square in this formula introduces systematic errors. However, we estimate that the
impact of $U/I$ to the total polarization ${\rm p}$ is less than $\pm 0.05~\%$.
Therefore we use ${\rm p}\cong Q/I$ and neglect the $U$ component in the subsequent discussion.

\subsection{The moonshine polarization}\label{s: moonshine}

As a check of our polarimetry we can compare the polarization
of the stray light from the moonshine with literature values from
\citet{Coyne1970} and \citet{Dollfus1971}. 

Forward scattering in the Earth atmosphere with scattering angles less than a few degrees does not introduce a
significant polarization effect. 
Therefore, we can assume that the polarization of the lunar stray light 
$(Q/I)_{\rm M}$ represents well the polarization of
the bright lunar crescent.
For areas just east of the Moon close to the focal mask (see Fig. \ref{cuts}) the scattered moonshine dominates
strongly. There we can neglect the contribution of the sky background $(Q/I)_{\rm sky}$ and assume that
\mbox{$(Q/I)_{\rm bg} \approx (Q/I)_{\rm M}$}. 

\begin{figure}
        \centering
        \includegraphics[width=8cm]{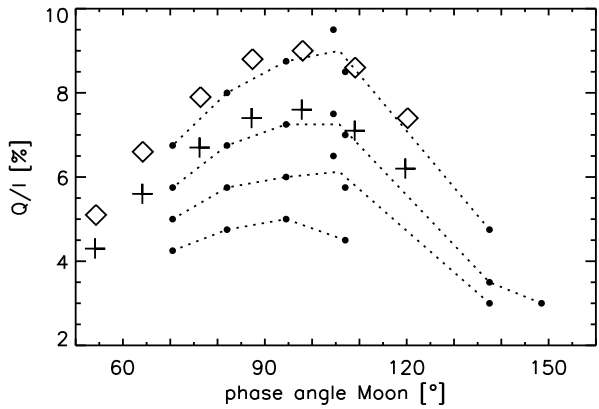}
                 \caption{Measured polarization of the lunar stray light near the focal mask
                 in the B, V, R, and I filters (from top to bottom by filled dots and dashed lines). 
                 Also indicated are the
                 polarization values given by \citet{Coyne1970} for the disk 
                 integrated moonshine
                 of the waxing Moon in their $B'$ ($\diamond$) and $G_m$ band ($+$).
                 }\label{moonshine}    
        \end{figure}

Figure \ref{moonshine} compares our results with
the waxing Moon values given by \citet{Coyne1970}. They used different
filters but their $B'$ and $G_{\rm m}$ bands 
($\lambda_{\rm eff}[\mu m]=0.45,\, 0.53$) are close to our B
and V band respectively and the good agreement with our data
underlines the consistency of our polarimetric data reduction.

Unfortunately \citet{Coyne1970} used no red filters
but the scaled $G_{\rm m}$ phase curve fits also well our 
R and I band data if scaling parameters of 0.80 and 0.65 are used respectively.
This is in good agreement with the wavelength dependency of the maximum degree
of polarization $P_{\rm max}$ of the whole Moon presented in
\citet{Dollfus1971}:
\begin{equation}  
P_{\rm max,\lambda_1}/P_{\rm max,\lambda_0}=(\lambda_1/\lambda_0)^{- 1.137} \,.
\end{equation} 
With this formula we obtain color ratios of
$q(R)/q(G_{\rm m})=0.81$ and $q(I)/q(G_{\rm m})=0.64$ for the Moon polarization in good agreement
with the above scaling parameters derived from our stray light data.

%=======================================================================

\section{A correction for the depolarization due to the 
back-scattering from the lunar surface }\label{s:retro-reflection}

The polarized light from the Earth is depolarized by the
back-scattering from the particulate surface of the Moon. 
We express this effect as polarization efficiency $\epsilon$,
describing the fraction of linear polarization preserved. 
We simplify the treatment by considering only the $Q$ linear 
polarization direction perpendicular and parallel to the
scattering plane Sun-Earth-Moon. Then the polarization
efficiency is
\begin{displaymath}
 \epsilon = {(Q/I)_{\rm es} \over (Q/I)_{\rm E}}\,,
\end{displaymath}
where $(Q/I)_{\rm E}$ is the Earth polarization. 
The polarization efficiency $\epsilon(\lambda,a_\lambda)$ depends on the wavelength and the
surface albedo. We neglect the phase dependence
in the back-scattering because the scattering angle is always
$179^\circ\pm 0.5^\circ$. 
 
The depolarization of the lunar surface  
was already investigated by \citet{Lyot1929} and \citet{Dollfus1957}. 
They measured the depolarization of the back-scattering 
of volcanic ashes and fines, which were used 
as proxy for the lunar soil. They found a well defined 
anticorrelation between albedo and polarimetric efficiency. 

Most important for the determination of $\epsilon(\lambda,a)$
are the albedo and polarization measurements for 
the reflection from several Apollo lunar soil
samples by \citet{Hapke1993, Hapke1998}. They illuminated 
eight samples under an inclination of 5~degrees (to avoid specular
reflection) with 100~\% polarized blue and red light and 
measured the ratio of $I_\perp/I_\parallel$ for phase angles $1^\circ$ 
($\sim$ back-scattering) to $19^\circ$ or scattering angles of 
$179^\circ$ to $161^\circ$. 
The results for the linear polarization ratio are presented in \citet{Hapke1993} in graphical form 
and we extracted the data for phase angle $1^\circ$ and derived 
the polarization efficiency $\epsilon$. The normal albedos are only
given for a phase angle of $5^{\circ}$ \citep[][Table~1]{Hapke1993} and we converted them into 
earthshine back-scattering
albedos corresponding to $1^{\circ}$ phase angle by applying a conversion factor of $1.25\pm 0.05$.
We derived this factor from albedo phase curves presented in \citet{Velikodsky2011} where they give a 
comprehensive summary of the results of
various independent photometric observations of the Moon including their own, Clementine data, and ROLO data.

       \begin{figure}
        \centering
        \includegraphics[width=8cm]{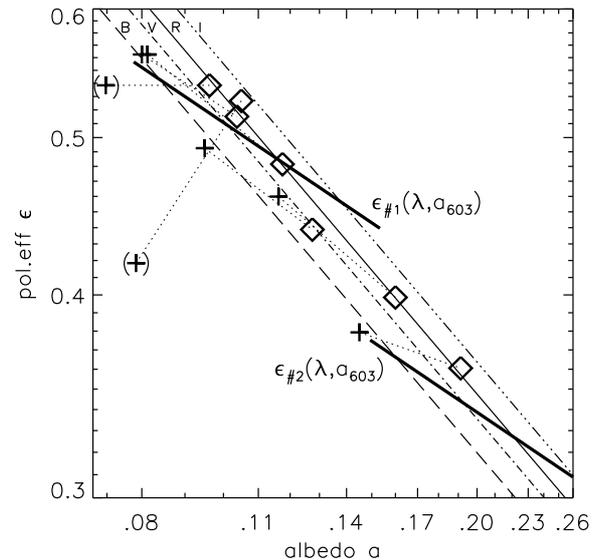}
                 \caption{Linear polarization efficiency as function of the normal albedo at $1^{\circ}$ (Eq. \ref{eq:pvsa}) for 
                  the B (dashed), V (dash-dot), R (solid), and I (dash-dotdot) band.
                  More information about the
                 fit procedure to  the \citet{Hapke1993} samples in the red ($\diamond$) and the blue ($+$) is given in the text.
                 Measurements of the same sample are connected by dotted lines.
                 The thick black lines show the derived wavelength
                 and albedo dependent polarization efficiencies for our two measurement areas \#1 and \#2 (Fig.~\ref{obs_areas}).
                 }
                 \label{pvsa}         
        \end{figure}

The samples from \citet{Hapke1993} include 5
low albedo samples $a_{\rm red}\approx 0.09-0.13$ representative 
for maria, 2 higher albedo samples $a_{\rm red}\approx 0.15 - 0.19$
representative for highlands and one non-typical, extremly high albedo 
sample with normal albedo $a_{\rm red}>0.35$. This sample with 
NASA number 61221 was taken from white material at the bottom of
a trench 
(see The Lunar Sample Compendium\footnote{http://curator.jsc.nasa.gov/lunar/compedium.cfm}) 
and therefore we treat this sample as special case in 
our analysis.

       \begin{figure*}
        \centering
        \includegraphics[width=18cm]{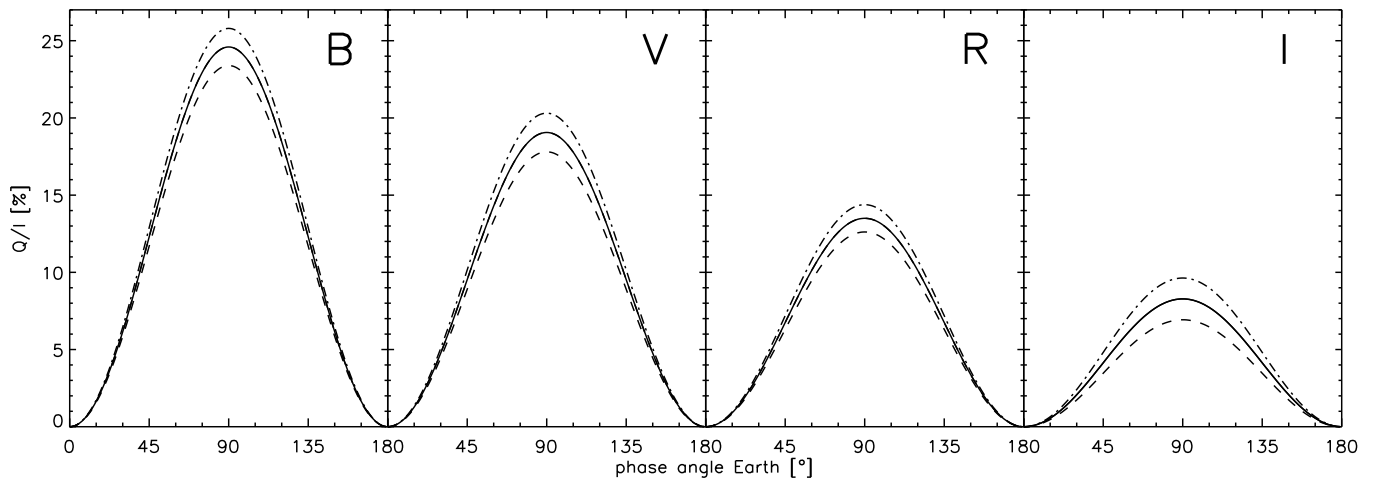}
                 \caption{Depolarization corrected polarization phase curves of Earth in the B, V, R, and I bands. 
                 The solid line indicates the mean of the mare (dashed) and highland (dash-dot) results
                 based on the polarization efficiency correction derived in this work. 
                 }
                 \label{results corrected}         
        \end{figure*}

Figure \ref{pvsa} shows the polarization efficiencies for the measurements
of \citet{Hapke1993} as function of the $1^\circ$-albedo for 
a blue wavelength ($\lambda=442$~nm) and a red wavelength 
($\lambda=633$~nm). 
The figure illustrates nicely the clear anticorrelation between
albedo $a$ and polarization efficiency $\epsilon$. 

We consider now in more detail the back-scattering properties of
the lunar samples. All samples, except 61221, show a very similar color 
dependence in their albedo with $a_{\rm red}/a_{\rm blue}= 1.35$ ($\sigma= 0.05$).
This is in agreement with the spectral variation of the
mean lunar albedo $\bar{a}$ from 
\citet{Dollfus1971} (see also \citealt{Gehrels1964}, Table XIII; and \citealt{Velikodsky2011}, Table 2)
described by 
\begin{equation}\label{albedo dependence}
{\rm log}\, \bar{a} =  0.83 \cdot {\rm log}\, \lambda [\mu{\rm m}] - 0.80\,.
\end{equation}
Inserting the wavelengths of the \citet{Hapke1993} measurements
into this formula yields 
$\bar{a}_{\rm red}/\bar{a}_{\rm blue} = 0.108/0.0805 = 1.34$
in very good agreement with the above derived albedo ratio for the lunar samples. 
 
For most samples the polarization efficiency is slightly higher 
(in one case equal) in the blue than in the red. 
There is one notable exception which is sample number 79221.
Although the albedo of sample 79221 is lower in the blue
than in the red its polarization efficiency is not higher
as in all other samples. Also when looking into reflectivity
studies for this sample \citep[e.g.][]{Noble2001}
it is not clear why this sample
could behave different in its depolarization properties 
than other maria soils. Therefore, we treat sample 79221 as
an exception.  
  
If we disregard sample 79221, then the remaining 6 samples 
have an average color dependence for their polarization 
efficiency ratio of  $\epsilon_{\rm red}/\epsilon_{\rm blue}= 0.91$ 
($\sigma = 0.05$). Including sample 79221 gives a mean ratio of
0.96 but a standard deviation which is with 0.14 significantly
higher.  

Based on these back-scattering measurements of lunar samples we
derive a two dimensional linear fit for the polarization
efficiency  ${\rm log}\,\epsilon$ as function of ${\rm log}\,a_{603}$, the albedo at 603 nm, and ${\rm log}\,\lambda$ for
the wavelength

\begin{equation}\label{eq:pvsa}
{\rm log}\, \epsilon(\lambda, a_{603}) = -0.61 \,{\rm log}\, a_{603}
- 0.291 \, {\rm log}\, \lambda [\mu{\rm m}] - 0.955 \,.
\end{equation}
For this fit the wavelength dependence of the albedo has been assumed to be according to Eq. \ref{albedo dependence} and it was normalized to 603 nm. 
By fitting the red data points we find the logarithmic slope $-0.61\pm 0.04$ between the polarization efficiency
$\epsilon$ and the \mbox{albedo $a_{\rm red}$} which we also adopt
for the blue data points. Finally, by fitting over the red and blue points separately we determine the other two parameters
$-0.291$ and $-0.955$. The resulting relation for the linear polarization efficiency as function of the normal albedo at $1^\circ$ is shown in
Fig. \ref{pvsa} for the B, V, R, and I band.

For the derivation of the albedos of our measurement regions we used the results of
\citet{Velikodsky2011} who present 
maps of lunar apparent and equigonal albedos at 
phase angles $1.7^{\circ}-73^{\circ}$ at wavelength \mbox{603 nm}. We extrapolated their results to a 
phase angle of $1^{\circ}$ and we get albedos 
$a_{\#1}(603~\rm nm)=0.11 \pm 0.01$ and $a_{\#2}(603~\rm nm)=0.21 \pm 0.01$.
The resulting polarization efficiencies
$\epsilon_{\#1}(\lambda,a_{603})$ and $\epsilon_{\#2}(\lambda,a_{603})$ are
listed in Table \ref{tbl: results} for the B, V, R, and I band and shown in \mbox{Figure \ref{pvsa}} giving the $\epsilon(\lambda, a_{603})$ 
fits.
Overall we estimate the uncertainty of the derived polarization efficiency to be
in the order of $\Delta\epsilon\approx \pm 3~\%$.

The main uncertainty of this derivation stems from the uncertainty in the above mentioned albedo conversion from $\alpha=5^\circ$ into albedos corresponding
to $1^\circ$ phase angle where the conversion factor $1.25\pm 0.05$ leads to an uncertainty of the polarization efficiency of about $\Delta\epsilon=\pm 1.5~\%$. 
To significantly improve the determination of $\epsilon$ accurate lunar albedo maps
for back-scattering geometry are required. This is because
for back-scattering at $\approx 1^\circ$
reflection is
strongly influenced by the opposition effect of the lunar surface, i.e. a steep brightness surge due to coherent backscattering and shadow-hiding
(e.g. \citealt{Shkuratov2011} and references therein). In addition to that the \citet{Hapke1993, Hapke1998} sample might not be representative for the surface properties of the Moon.

The logarithmic slope $-0.61\pm 0.04$ is better constraint for the low albedo samples and it introduces an uncertainty  $\Delta\epsilon = \pm 1~\%$ towards the higher albedo samples.
Moreover, the logarithmic relation might not be valid
over the complete albedo range between $a_{\lambda}=0.09-0.19$ and two slopes, one for the maria and one for the highlands, might be necessary. However, based on the
available samples this is not obvious and one log fit may not be the best representation of the data. 
More direct measurements of the polarization efficiency of the lunar back-scattering are required to reduce this source of uncertainty.

%=======================================================================

\section{Polarization of planet Earth} \label{s: earth polarization}

\subsection{Fractional polarization derived from the earthshine}

In Figure \ref{results corrected} we present the depolarization-corrected polarization
phase curves of planet Earth
in the B, V, R, and I bands and the corresponding corrected fit parameters $q_{\rm max,corr}$ are listed in Table \ref{tbl: results}. 
For the B band we obtain a maximum polarization of about $25~\%$ which decreases with wavelength to about $8~\%$
in the I band. For perfect measurements and perfect polarization efficiency corrections the same Earth polarization $q_{\rm max,corr}$
values should be obtained for the mare and highland regions. We note that the corrected highland results are systematically
higher than the mare results by a factor of about $1.1$ for the B, V, R bands and 1.4 for the I band. This reflects also the uncertainty in our determination of the
polarization efficiency of the back-scattering $\Delta \epsilon \approx \pm 3~\%$ derived in Section \ref{s:retro-reflection}.

\subsection{Comparison with previous measurements}

     \begin{figure}
        \centering
        \includegraphics[width=8cm]{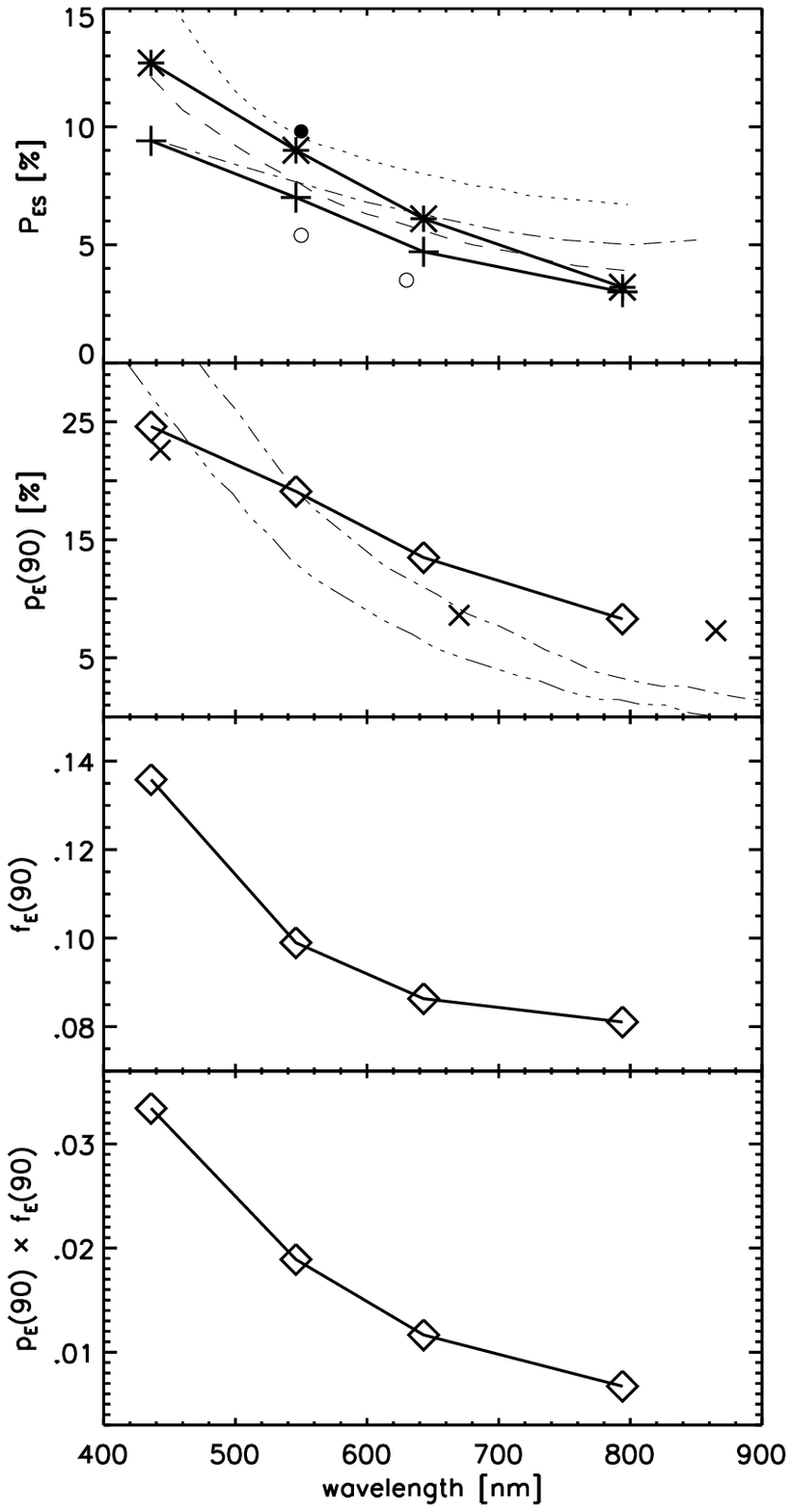}
                 \caption{Top: Earthshine polarization results at quadrature for maria ($\ast$) and highlands ($+$).
                 The thin lines give the \citet{Sterzik2012} spectro-polarimetry for waning (dashed) and waxing (dotted)
                 moon at Earth phases $87^\circ$ and $102^\circ$ respectively and the \citet{Takahashi2013} spectro-polarimetry (dash-dot) at $96^\circ$. 
                 The circles are the \citet{Dollfus1957}
                 values $q_{\rm max}$ from Fig. \ref{results} (filled) and two additional observations at
                 Earth phase \mbox{$\approx100^\circ$} (open). $2^{nd}$ panel: Earth polarization $p_{\rm E}$ from Table \ref{tbl: quadrature results} 
                 ($\diamond$) compared to the 
                 POLDER/ADEOS results of \citet{Wolstencroft2005} ($\times$) and two \citet{Stam2008} models with
                  40~\% (dash-dot) and 60~\% (dash-dotdot) cloud coverage.
                 Bottom two panels: spectral reflectivity of Earth $f_{\rm E}$ and
                 polarized reflectivity of Earth $p_{\rm E}\times f_{\rm E}$.}
                 \label{ColorDep}         
        \end{figure}

In Fig.~\ref{results} we compare our earthshine 
measurements with the pioneering
study of \citet{Dollfus1957}, who obtained his data with visual
observations using a ``fringed-field polariscope''. The agreement with
our V band phase curve for the mare region is excellent. If we apply
$q_{\rm max} sin^2$ fits (see Sect. \ref{phase dependence}) to both data sets  
the quadrature signals only differ by
0.8~\%. 
The small deviations between \citet{Dollfus1957} and us can be explained
by different mare regions observed and the expectedly non-equal effective wavelengths of the two
completely different types of measurements.
For one night at $\alpha\approx 100^\circ$ \citet{Dollfus1957} reports also 
the earthshine polarization in two filters, namely $p=5.4~\%$ for 0.55 $\mu$m (V' band)
and 3.5~\% for 0.63 $\mu$m (R' band). The ratio $p_{V'}/p_{R'}= 1.54$ is again in
excellent agreement with our
polarization ratio $q_{\rm max,V}/q_{\rm max,R}=1.47$.
This indicates that the filters used by \citet{Dollfus1957} must match quite well
our filter pass bands. 
  
The spectral dependence of the earthshine polarization observed with a spectral resolution
of 3~nm was recently
published by \citet{Sterzik2012}. These sensitive spectro-polarimetric data reveal weak,
narrow features of the planet Earth due to O$_2$ and H$_2$O on 
a smooth polarization spectrum decreasing steadily from the blue towards longer wavelengths. 
They present measurements for two epochs
with phase angles $\alpha=87^\circ$ for a waning moon phase 
and $\alpha=102^\circ$ for the waxing moon phase. 
For the waning
moon case they obtained an earthshine polarization of about $p_{\rm B}=12.1$~\% in 
the B band, $p_{\rm V}=7.7$~\% in V, $p_{\rm R}=5.6$~\% in R, and $p_{\rm I}=3.9$~\% in I, and 
a significantly higher polarization for the waxing moon phase with $p_{\rm B}=16.6$~\%, 
$p_{\rm V}=9.7$~\%, $p_{\rm R}=8.0$~\%, $p_{\rm I}=6.7$~\% as plotted in Fig. \ref{ColorDep}.
Unfortunately it is not clear whether they measured 
the back-scattering from maria or highlands.
\citet{Sterzik2012} attribute
the polarization differences between the two epochs mainly to intrinsic differences 
of the polarization of Earth because the earthshine stems from 
different surface areas and were taken for days with different cloud coverage. 
Considering our polarization values for highlands and
maria then it could be possible that the differences
measured by \citet{Sterzik2012} are at least partly due to the mare/highland depolarization difference 
(or surface albedo difference).

Another spectra-polarimetric observation of the earthshine was published by \citet{Takahashi2013}. 
They also find a rise of the fractional polarization of the earthshine towards the blue but with a much flatter slope.
Unfortunately they do not report whether their results were obtained from maria or highlands either. Therefore, only a qualitative
comparison with our data can be made. The observations of \citet{Takahashi2013} are conducted at 5 consecutive nights
and cover phase angles $\alpha=49^\circ - 96^\circ$. In the blue they find that the maximum polarization
is reached at $\alpha \approx 90^\circ$. However, for wavelengths $> 600~{\rm nm}$ 
the polarization keeps increasing up to and including their last measurement at $\alpha=96^\circ$.
They conclude that the phase with the highest fractional polarization $\alpha_{\rm max}$
is shifted towards larger phase angles which could be explained by an increasing contribution of the
Earth surface reflection.
In our data we do not see this shift but neither can we exclude it because 
we were not able to derive meaningful data due to the very strong stray light from the moonshine and the weak signal from
the earthshine.
In this regime our linear extrapolation method to subtract
the background stray light from the earthshine signal introduces a strong systematic overestimate $\Delta_{\rm syst}$ of the result (see Sect. \ref{s: extract earthshine}).
\citet{Takahashi2013} also use a linear extrapolation method to determine the earthshine polarization but unfortunately they do not describe their data reduction in detail.
Therefore, considering the limitations of our linear extrapolation, it could be possible that the shift of  $\alpha_{\rm max}$ reported by \citet{Takahashi2013}
is due to the strong stray light  at phase angles $> 90^\circ$.

Overall, the spectral dependence of the polarization of \citet{Sterzik2012} and \citet{Takahashi2013} is qualitatively
similar to our measurements but the level and slope of the fractional polarization differ quantitatively. 
Because \citet{Sterzik2012} and \citet{Takahashi2013} provide no information about the lunar surface albedo for their
measuring area and do not assess the stray light effects from the bright moonshine their results cannot be used for a
quantitative test of our results.
The spectral slope of \citet{Sterzik2012} is slightly steeper than ours while the slope of \citet{Takahashi2013} is slightly flatter.

For an assessment of the polarization efficiency for the lunar back-scattering we used 
literature data for polarimetric
measurements of lunar samples by \citet{Hapke1993, Hapke1998} and we derive
a wavelength and surface albedo dependent polarization 
efficiency relation $\epsilon(\lambda,a_{603})$ 
which gives for mare in the V band
 $\epsilon(V,0.11)=50.8~\%$. This value is
significantly higher than the 33~\% derived by 
\citet{Dollfus1957} which he based on the analysis of volcanic
samples from Earth used as a proxy for the lunar maria. 
Because of this, the Earth polarization derived in this
work is much lower than the value given in \citet{Dollfus1957}. 
We are not aware of other studies on the polarization efficiency $\epsilon$
for the lunar back-scattering. Relying the determination of $\epsilon$ on
real lunar soil is certainly an important step in the right direction for a more accurate
determination of the polarization of Earth.

Very valuable are the reported Earth polarization values
from \citet{Wolstencroft2005} based on direct
polarization measurements with the POLDER instrument on the
ADEOS satellite. They derived the fractional polarization for the
wavelengths 443 nm (B'), 670 nm (R') and 865 nm (I') for different surface types and
cloud coverage. 
Weighted mean values representative for an integrated planet 
Earth observation (55~\% cloud coverage) 
of 22.6~\%, 8.6~\% and 7.3~\% in the 
B', R' and I' band are obtained which are also indicated in the second panel of
Fig.~\ref{ColorDep}. The good agreement between our 
derivation based on the earthshine and the values from 
\citet{Wolstencroft2005}
confirms our determination of the polarization efficiency. 
Unfortunately, it is not possible to assess whether the R band point of this study
differs significantly from the value of \citet{Wolstencroft2005}
because they give no description of their data and uncertainties.

\subsection{Comparison with the models from Stam (2008)}

The study of \citet{Stam2008} is unique for the modeling of
the spectral dependence of the fractional polarization of
Earth-like planets. In her work she explored also dependencies on a
range of physical properties different from Earth. For our
comparison we pick the model for an
inhomogeneous Earth-like planet with 70~\% of the surface
covered by a specular reflecting ocean and 30~\% by deciduous 
forrest (lambertian reflector with an albedo for forest), and cloud coverages 40~\% and 60~\%. When compared to our
Earth polarization determinations (Fig. \ref{ColorDep}, second panel), these models
agree with our measurements at short wavelengths but
show a clear deficit in the fractional polarization at long wavelengths in the I band.
This is not surprising
since the models were not tuned to the case of Earth. In the models only very thick, liquid water clouds were included but no thin liquid water clouds and no ice clouds.
\citet{Karalidi2012} showed that with more realistic cloud properties for Earth the degree of polarization can vary strongly depending on cloud optical thickness.
Hence, our data could now be used to test and to improve model calculations for the Earth polarization.

\subsection{Polarization flux contrast for the Earth - Sun system} \label{flux contrast}

   \begin{table}
        \caption{Geometric albedo $A_g$, phase integral $A_s/A_g$ and quadrature results for the planet Earth. 
        		    Given are the fractional polarization $p_{\rm E}$, the Earth reflectivity $f_{\rm E}$,
        		     the polarized reflectivity \mbox{$p_{\rm E}\times f_{\rm E}$}, and the polarization contrast 
		     $C_{\rm p}=p_{\rm E} \times f_{\rm E} \times \left(R_{\rm E}/d_{\rm S-E}\right)^2$ for an Earth-Sun system.
		     The systematic offset uncertainty $\Delta p_{\rm E}$ is due to the uncertainty of the 
		     depolarization and albedo of the lunar surface.
                       }                               
        \label{tbl: quadrature results}      
        \centering          
                 \begin{tabular}{L{25mm} | C{10mm} C{10mm} C{10mm} C{10mm}}    
            \hline   
                                              &        B         &        V           &      R           &         I         \\
            \hline                
            $A_g$      &    0.504       &     $0.367^1$          &     0.320      &         0.301    \\
            $A_s/A_g$      &    0.86       &     0.86          &     0.86      &         0.86    \\
                      \hline
                           values for $\alpha = 90^\circ$                       &                &                  &                 &            \\
            \hline                
            $p_{\rm E}$ [\%]      &    24.6       &     19.1          &     13.5      &         8.3      \\
            $\Delta p_{\rm E}$ [\%]                      &    $\pm 1.2$       &   $\pm 1.3$         &  $\pm 0.9$     &   $\pm 1.4$   \\
            $f_{\rm E}$                                               &    0.136       &   0.099         &  0.086     &   0.081     \\
            $p_{\rm E} \times f_{\rm E}$   [$10^{-3}$]              &    33.41       &   18.90         &  11.66     &   6.73       \\
            $C_{\rm p}$ [$10^{-11}$]            &    6.07      &   3.44        &  2.12     &   1.22       \\
            \hline

            \multicolumn{5}{l}{\vspace{-0.2cm}}\\
             \multicolumn{5}{l}{$^1$\citet{Cox2000}}
               \end{tabular}
        \end{table}

      \begin{figure}
        \centering
        \includegraphics[width=8cm]{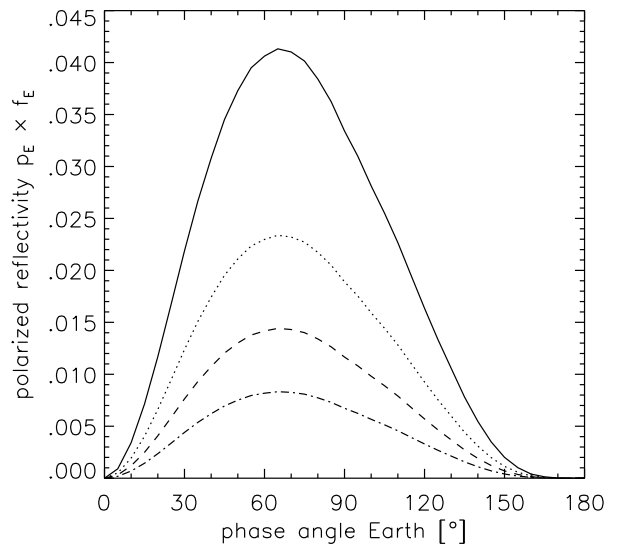}
                 \caption{Polarized reflectivity phase curve $p_{\rm E}(\alpha)\times f_{\rm E}(\alpha)$ for Earth in the B (solid), V (dotted), R (dashed), and I band (dash-dot).}
                 \label{phasefunction}         
        \end{figure}

A key parameter for the polarimetric search and characterization of Earth-like 
extra-solar planets is the polarization flux $p \times f$ of a planet or the
polarization flux contrast $C_p$ as described in \mbox{Eq. \ref{cpol}}. 
The polarization flux of a highly polarized planet is easier to 
measure than the fractional polarization $p$, because the reflected intensity
cannot be distinguished easily from the scattered light halo of the central
star in high contrast observations. But because stars are essentially unpolarized, it should be possible
to detect a differential signal of polarized light from extra-solar planets
with high contrast polarimetric imagers as foreseen for the upcoming instrument 
SPHERE/VLT and planed for future facilities like the \mbox{E-ELT}
\citep[e.g.][]{Schmid2006, Beuzit2008, Kasper2010}.
The polarimetric detection of an Earth-like planet is difficult but, nonetheless, 
it is useful to have accurate values for the expected signal for the planning
of such observations.

The prediction of the polarization flux of an exo-Earth requires
besides the fractional polarization $p(\alpha,\lambda)$ determined in this 
work also the reflected intensity $f(\alpha,\lambda)$. 

The reflected intensity of Earth can be split into a wavelength
dependent geometric albedo term $A_g(\lambda)=f(0^\circ,\lambda)$
and a normalized phase dependence $\Phi(\alpha)/\Phi_0$ 
where $\Phi_0=\Phi(\alpha=0)$ according to:
\begin{displaymath}
f(\alpha,\lambda) = A_g(\lambda) {\Phi(\alpha)\over \Phi_0}\,.
\end{displaymath}
With this approach we neglect the color dependence of the phase curve, which 
is not known but certainly small when compared to the measuring
uncertainties for the spectral albedo $A_g (\lambda)$ and the uncertainties in the fractional
polarization $p_{\rm E}(\lambda, \alpha)$. 
  
The visual geometric albedo of Earth is $A_g(V)=0.367$ \citep{Cox2000}.
With the relative spectral reflectance 
measured by \citet{Arnold2002}, \citet{Woolf2002}
and 
\citet{Montanes2005}, we deduce the geometric albedo for the
individual B, V, R, and I filters as given in \mbox{Table \ref{tbl: quadrature results}}.

For the phase dependence $\Phi(\alpha)/\Phi_0$ we use   
the phase curve determined by \citet{Palle2003} for the 400-700~nm filter
normalized to the B, V, R, and I band geometric albedos derived above. The derived
phase curves are given in Fig. \ref{evsm} and 
their value for $\Phi(90^\circ)/\Phi_0 =  0.27$ yields  
the polarized reflectivity for quadrature phase $p_{\rm E}(90)\times f_{\rm E}(90)$
for the B, V, R, and I filters as plotted in Fig. \ref{ColorDep}
and given in Table \ref{tbl: quadrature results}. 

The phase curve $\bar{f}_{\rm E}$ of \citet{Palle2003} is 
based on earthshine measurements
at phases between $\alpha = 30^\circ - 145^\circ$ extrapolated to $\alpha = 0^\circ - 180^\circ$. 
This broad phase angle coverage is unique and remains, to our knowledge, the only available observation of the 
phase dependence $\Phi(\alpha)$ of $f_{\rm E}$. For future reference we also give in Table \ref{tbl: quadrature results} the 
phase integral parameter $A_s/A_g$, the ratio between spherical and geometric albedo, derived from the 
\citet{Palle2003} data.

The spectral dependence of the polarization flux $p_{\rm E}(\lambda, 90^\circ) \times f_{\rm E}(\lambda, 90^\circ)$ 
of Earth decreases steeply towards longer wavelength, because
both, the fractional polarization $p_{\rm E}$ and the reflectivity $f_{\rm E}$ are
higher for the blue than the red. The $p_{\rm E} \times f_{\rm E}$ signal in the B band is about a 
factor 5 times stronger than in the I band.  

The polarization contrast $C_p(\lambda,90^\circ)$ according to Eq. \ref{cpol}
is determined from $p_{\rm E} \times f_{\rm E}$ and using $R_E^2/{\rm AU}^2$. This yields values
at the level of a few times $10^{-11}$ only 
(Table \ref{tbl: quadrature results}). One should note that
an Earth-like planet in the habitable zone of a M star with
$L=0.02~L_\odot$ is at a much smaller separation of 0.14 AU. In this
case the expected polarization contrast is about a factor of 50 higher
and within reach for a high contrast imaging polarimeter at
an ELT (Kasper et al. 2010).

The B, V, R, and I phase curves for the polarized reflectivity 
\mbox{$p_{\rm E} \times f_{\rm E}$} are plotted in Fig. \ref{phasefunction}.  
The maximum signal occurs near $\alpha\approx 65^\circ$, which is thus the 
best phase for a detection.

\section{Summary and Discussion}\label{s:conclusions}

This work presents measurements of the earthshine polarization in the
B, V, R and I bands. The data were acquired with a specially designed
wide field imaging polarimeter using a focal plane mask for the suppression
of the light from the bright lunar crescent. Thanks to this measuring
method we can accurately correct for contributions from the (twilight) sky and 
the scattered light from the bright lunar crescent and derive
values with well understood uncertainties. We derive phase curves
for the fractional polarization for the earthshine reflected from maria
and highlands for the different filter bands. The phase curves can be
fit with the sine-square function $q_{\rm max} \sin^2(\alpha)$. The amplitude
$q_{\rm max}$ decreases strongly with wavelength from about 
13~\% in the B band to about 3~\% in the I band (see Table \ref{tbl: results}). The fractional 
polarization of the earthshine is about a factor 1.3 higher for 
the dark mare region when compared to the bright highland. 
Our phase curve for the mare region in the V band is in very good agreement with 
the historic visual polarization phase curve from \citet{Dollfus1957}.

We study the depolarization introduced by the back-scattering from
the lunar surface based on published polarimetric
measurements of lunar samples \citep{Hapke1993, Hapke1998}. 
We derive a 2-dimensional fit function for the polarization efficiency 
$\epsilon(\lambda,a_{603})$ for the back-scattering 
which depends on wavelength and surface albedo. 
Earthshine measurements plus $\epsilon$ correction yield
as main result of this paper the fractional polarization of the 
reflected light from the planet Earth as function of phase in four bands. 
The polarization of Earth at quadrature phase is as high as 
25~\% in the B band and decreases steadily with wavelength to 8~\% in the
I band (see Table \ref{tbl: quadrature results}). 
Similar values were reported from direct satellite measurements
of the Earth polarization \citep{Wolstencroft2005}.  

This work provides the most comprehensive measurements
of the polarization of the integrated light of the planet Earth up to now.
The determined values can be used as benchmark values
for tests of polarization models and for predictions for future polarimetric 
observations of Earth-like extra-solar planets. 
In particular we describe accurately our measurements and assess the uncertainties.
In addition we apply for the first time a polarization efficiency $\epsilon$ correction
which is based on lunar soil measurements, and which is significantly different from 
previously used volcanic rock measurements.

Similar to our data of Earth the models of \citet{Stam2008} for horizontally 
inhomogeneous Earth-like planets with thick liquid water cloud coverage show 
also a decrease in fractional
polarization with wavelength but with a significantly steeper slope.  
This may indicate that other
scattering components, for example aerosols, thin liquid water clouds, and ice clouds contribute
significantly to the Earth polarization in the I band \citep[see][]{Karalidi2012}.

Are our polarization values for the planet Earth representative
or should we expect large temporal variations? Our data were
taken during two observing runs lasting each a few days.
Two data sets are from similar phase angles,
$73.0^\circ$ and $75.5^\circ$, but they were taken 7 months apart. 
The measured fractional polarization differs by about $\Delta q/q\approx 0.1$. 
Also the deviation of the data points from the fit $q_{\rm fit}=q_{\rm max}\sin^2\alpha$ is 
at the same level $(|q-q_{\rm fit}|)/q\approx 0.1$. This scatter is at the level 
of our calibration errors. Therefore, variation of the intrinsic polarization signal of
Earth on the 10~\% level could be present in our data without
being recognized. Our measurements show certainly no changes at
the $\Delta q/q\approx 0.3$ level as suggested by \citet{Sterzik2012}.
Variations in the fractional polarization are of interest because they
could be used as diagnostic tool for investigations of surface structures
or temporal changes in the cloud coverage of extra-solar planets. 

Because our study includes a detailed assessment of the 
uncertainties for each step in our determination, we can now discuss 
how the Earth polarization measurement could be improved.

Polarization variability studies could be carried out
with enhanced sensitivity selecting observing periods and filters
with strong earthshine polarization signals in order to minimize
statistical noise and systematic effects in the data extraction. 
Observations in the B and V filter, and for phase
angles in the range from $\alpha=40^\circ$ to $100^\circ$ would be
ideal for such studies. Measurements taken for several
consecutive nights would allow a sensitive search for day to day
variation at a level of 
$\Delta q/q \approx 0.03$ due to variable cloud 
coverage. Also multiple epoch data could be collected for an investigation 
of long term and seasonal polarization changes. 

The determination of a more accurate wavelength dependence of the 
earthshine polarization could be established with long integrations
for phase angles between $50^\circ - 80^\circ$ when the earthshine
polarization signal is strong, the level of scattered light from the
moonshine still low, and the time for observations after twilight long
enough for observations in multiple filters. 

More accurate phase curves require a careful analysis of the data
from different phases because the earthshine observing conditions
and the associated measuring and calibration procedures change
strongly with lunar phase. If these problems can be 
solved then one could determine accurately the peak in the fractional 
polarization curve near $\alpha = 90^\circ$ as function of wavelength
and investigate the presence of a rainbow feature in the polarization data
around $\alpha=40^\circ$ \citep[see][]{Stam2008}. 

A more accurate absolute value for the polarization of the planet Earth
requires first more data in order to average out intrinsic 
variations. Equally important is a more accurate determination of
the surface albedo for the measuring region and the associated
polarization efficiency $\epsilon(\lambda,a_\lambda)$ for the correction
of the lunar back-scattering. 

The imaging polarimetry of the earthshine presented in this study
and the spectro-polarimetric results from \citet{Sterzik2012} and \citet{Takahashi2013} demonstrate
that the investigation of the Earth polarization via earthshine measurements 
is very useful and attractive. Detailed and versatile investigations 
are possible with existing polarimetric instruments as used by \citet{Sterzik2012} and \citet{Takahashi2013}
or with small, specific experiments as demonstrated in this work. 
The obtained results can be compared with model calculations like
those described in \citet{Stam2008} and teach us about light scattering
processes of planets. Because we know so well our Earth we can also
investigate subtle effects, which are potentially important in other
planets. Building up our knowledge on scattering polarization from Earth
could therefore become also important for the future polarimetric
investigation of extra-solar planets.

%=======================================================================
  
\begin{acknowledgements}
      Part of this work was supported by the FINES research fund by a grant through the Swiss National Science Foundation (SNF).
\end{acknowledgements}

%=======================================================================

%%%%%%%%%%%%%%%%%%%%%%%%%%%%%%%%%%%%%%%%%%%%%%%%%%%%%%%%%%%%%%%%%%%%%%%%%%%%
%% references
\bibliographystyle{aa}    %% aa.bst = A&A bibliography style
\bibliography{earthshine}    %% example.bib = Bibtex entries from ADS

\end{document}